# Competing Mechanisms between Dislocation and Phase Transformation in Plastic Deformation of Single Crystalline Yttria-Stabilized Tetragonal Zirconia Nanopillars

Ning Zhang and Mohsen Asle Zaeem[*]

Department of Materials Science and Engineering, Missouri University of Science and Technology, Rolla, MO 65409 USA

**Abstract**

Molecular dynamics (MD) is employed to investigate the plastic deformation mechanisms of single crystalline yttria-stabilized tetragonal zirconia (YSTZ) nanopillars under uniaxial compression. Simulation results show that the nanoscale plastic deformation of YSTZ is strongly dependent on the crystallographic orientation of zirconia nanopillars. For the first time, the experimental explored tetragonal to monoclinic phase transformation is reproduced by MD simulations in some particular loading directions. Three distinct mechanisms of dislocation, phase transformation, and a combination of dislocation and phase transformation are identified when applying compressive loading along different directions. The strength of zirconia nanopillars exhibits a sensitive behavior depending on the failure mechanisms, such that the dislocation-mediated deformation leads to the lowest strength, while the phase transformation-dominated deformation results in the highest strength.

---
[*] Corresponding author.
   Email address: zaeem@mst.edu

1. Introduction

Zirconia (ZrO$_2$) has three polymorphs, i.e., monoclinic (*m*), tetragonal (*t*) and cubic (*c*) phases. At room temperature, only the low symmetry monoclinic phase is thermodynamically stable, and around 1480 K, the first-order martensitic transition to the tetragonal phase would be triggered, and further converted to the higher symmetry cubic fluorite phase at 2573 K [1]. Prior to 1970s, pure monoclinic zirconia was of very limited interest due to the crumbling (e.g. fracture, crack) of the ceramic components commonly observed during cooling from the tetragonal phase, which is accompanied by a volume expansion of ~4% [2]. In contrast, the high-temperature phases of zirconia, i.e., tetragonal and cubic, have excellent mechanical, thermal, chemical and dielectric properties including high-strength, low thermal conductivity, high corrosion resistant, and high ionic conductivity, which lead zirconia to a wide range of industrial, technological and medical applications. Currently zirconia is employed as, for example, catalytic support medium [3], electrolyte in solid oxide fuel cells at low temperature [4, 5], thermal barrier coating [6], nuclear waste confinement [7], one of the leading candidates for alternative gate dielectrics [8], and in shape memory [9, 10] and dental applications [11].

In general, the stabilization of high symmetry polymorphous of zirconia, i.e., *t*- and *c*-ZrO$_2$, at room temperature can be achieved by doping a variety of oxide additions [12], one of which is yttria (Y$_2$O$_3$), or by reducing the grain size to nanometer scale [13, 14]. The *t*-ZrO$_2$ grains have been observed to be size-stabilized below 200nm [15], while *c*-ZrO$_2$ was stated to be stable at a few nanometers scale [16].

The discovery, that the martensitic tetragonal-to-monoclinic (*t* → *m*) transformation in zirconia can be controlled to serve as the source of transformation plasticity, heralded new visions for the high-performance applications of zirconia [2]. Transformation toughening of zirconia was first reported by Garvie, Hannink and Pascoe in 1975 [17]. Yttria-stabilized tetragonal zirconia (YSTZ) presents a stress-induced phase transformation of tetragonal to stable monoclinic form, which is accompanied by a volumetric expansion to close crack tips and superimposes compressive stresses on the existing stress field, hence, makes the material more resistant to crack propagation [17]. During the past decades, considerable efforts have been devoted to study $t \rightarrow m$ phase transformation behavior of stabilized zirconia. This transition not only is of intrinsic interest, but is also important because it has been suggested that the reverse transition, i.e., from the monoclinic to tetragonal phase, should be related to the recognition of the potential shape memory and superelastic ceramics [9, 10].

Theoretically, transformation-induced plasticity and transformation toughening require a reliable determination of the strain field in the stress-activated transformation zone [2], and this, in turn, requires a detailed understanding of transformation crystallography. Thus, several phenomenological theories have been developed based on the crystallographic characteristics. For example, the crystallographic theory [18], which describes the structural change by a homogeneous lattice deformation, was applied to calculate the strains associated with formation of individual units of monoclinic product in a tetragonal matrix. In addition, a generalized theory of martensite crystallography [19, 20] was employed to provide the quantitative data that is essential for developing a credible, comprehensive understanding of the transformation toughening process.

Experimentally, the $t \rightarrow m$ phase transformation has been studied mainly by optical methods or transmission electron microscopy (TEM) [21]. Both the positive (transformation toughening) and negative consequences (low temperature degradation, microcracking) of this transformation have been investigated. By annealing in water, the polycrystalline yttria-doped tetragonal ZrO$_2$ on the surface of the sintered body was observed to transform to the monoclinic phase, accompanied by microcracking [22]. In addition, using Raman spectroscopy, atomic force microscopy (AFM) and optical microscopy, catastrophic cracks were detected during phase transformation in zirconia films due to the change of circumferential stress. In turns, such cracks were found to lead to breakaway oxidation of zircaloy [23]. The phase composition and structure of zirconia single crystal doped with Y$_2$O$_3$ was studied by X-ray diffraction analysis and TEM, in which twinning hierarchy was observed to facilitate the elastic stress relaxation [1]. On the other hand, under high-pressure torsion a $t \rightarrow m$ phase transformation with a coherent interface occurred. Once the monoclinic phase reached the saturation level, nanograins of high dislocation density but with no twins were formed [24]. With a small amount of Al$_2$O$_3$ doped in 3.0 mol% Y$_2$O$_3$-stabilized tetragonal zirconia polycrystal, the cubic phase regions were formed and the grain-growth rate was enhanced by the diffusion-enhanced effect of Al$_2$O$_3$-doping [25]. It is now generally recognized that the size of grains has a strong effect on the transformation toughening. Alternatively, fracture toughness of nanocrystalline zirconia was measured by nano-indentation methods, and it was found that the larger the tetragonal grains the greater propensity to undergo martensitic transformation into stable monoclinic structure, consequently enhance the toughness of material [26]. Recently, to overcome the shortcomings of bulk zirconia ceramics that tend to crack when undergo martensitic transition, a fine-scale structure with few crystal grains was fabricated. Such oligo crystalline structures were found to be capable of superelastic cycles and robust shape memory by virtue of reduction of internal mismatch stress during phase transition [9, 10, 27].

Recently, computer simulations have become uniquely powerful tools for addressing issues of phase transition at atomistic and micro scales. Indeed, important insights have already come from simulations; for example, first–principle [28] and *Ab initio* calculations [29] have been performed to look into the structural and electronic properties of pure zirconium and yttria-stabilized cubic zirconia. More recently, a two-dimensional elastic phase field model was proposed with an attempt to study the effect of external stress on martensitic phase transformation and transformation toughening. Such model was demonstrated to be capable of predicting the transformation zone and stress field around crack tips [30], as well as the pseudoelasticity behavior exhibits in shape memory application [31]. Following on, three-dimensional phase field modelling was carried out to investigate the effect of martensitic variant strain accommodation on the microstructural formation and topological patterning in zirconia [32]. Despite the tremendous capabilities of phase field modelling in predicting microstructure evolutions at the mesoscale, it does not explicitly deal with the behavior of the individual atoms, which is essentially importance for an in-depth understanding of the physics of phase transformation in zirconia [33]. Moreover, hybrid Monte Carlo-molecular dynamics simulations [34] were carried out to study defect distributions near tilt grain boundaries in nanocrystalline YSTZ. Very recently, a new approach that adapted the first-principle model has been reported to analysis the superplastic behavior of fine-grained YSTZ polycrystal [35]. With the advantages of offering tight control on composition and microstructure, as well providing valuable atomistic information of structure evolutions, molecular dynamics (MD) technique has been widely used as a powerful computational tool for studying the mechanisms of stabilization [36], thermal transport [37-40], diffusion of oxygen ions [41-44], and cubic-to-tetragonal phase transition [45] of zirconia. However, MD simulations of YSTZ are by now mainly concentrated on polycrystalline structure and the cubic polymorph of zirconia. To understand how the texture affects the mechanical properties, one should first examine mechanical properties of single crystals and interpret the deformation processes involved. However, the fundamental understanding of *t→m* transformation behavior at atomic level is less investigated up to this time, and to the best of our knowledge and despite its significance, $t \to m$ phase transition in single crystalline YSTZ has not been investigated so far.

In addition to $t \to m$ martensitic phase transformation, other microstructural features, such as dislocation, can also play a pivotal role in determining the mechanical properties of material. Dislocation migration, as one of the dominated deformation mechanisms of cubic zirconia, has been extensively studied. And dislocations were found to preferentially occur on cube planes, such as {100}, {110} and {111} planes [46]. For tetragonal zirconia, a ferroelastic behavior preceding dislocation plasticity [46] was reported experimentally. Under tension loading, a tetragonal single crystal forms containing residual defects. Although the dislocation pile-ups activity at the interface between grains in nanocrystalline cubic zirconia [47, 48] are receiving much attention, there are relatively fewer studies on dislocation behavior of single crystalline tetragonal zirconia, especially under compressive loading.

For size-limited single crystals, as one traverses to the nanometer scale, phase transition and dislocation nucleation behavior becomes ultra-difficult to capture for the present *in situ* nanoscale tomography compared to the bulk parent material. In addition, the accurate orientation control of material, which is particularly important for anisotropic crystals, arises another challenging issue for experiments. For example, in a very recently work by Zeng et al. [49], a study on martensite mechanics was conducted by using micro-scale YSTZ specimens; the results indicated that the stress-induced martensitic transformation in zirconia-based shape memory ceramics depends on the orientation of tetragonal crystals. However, the detailed evolution of phase transformation and dislocation motion during plastic deformation is very hard to capture by the current experimental techniques. Apparently, a lack of ground understanding of the atomistic deformation mechanisms in materials severely limits our ability to design nanomaterials with desired mechanical properties. Therefore, atomistic computer simulation, which has been demonstrated to be successful in modeling phase transition and dislocation behavior regarding orientation effect under compression [50, 51], is ideally suited to provide valuable atomistic information of zirconia.

In this study, MD simulation is applied to the single crystalline yttria-stabilized tetragonal zirconia (YSTZ) nanopillars to investigate the behavior of $t \rightarrow m$ phase transformation and dislocation migration under uniaxial compression. The aim of the study is to systematically explore the atomistic deformation mechanisms of single crystalline tetragonal zirconia with respect to different crystallographic orientations.

This paper is organized as follows. Section 2 describes the details of the computational approach and atomistic models. Section 3 presents the simulation results with full discussions regarding orientation effect on plastic deformation mechanisms. Section 4 summarizes our conclusions.

## 2. Computational Details

### 2.1 Interatomic potential

Selecting a suitable interatomic potential is a crucial step for any MD simulation. The potential energy is a function of the distance between ions. For YSTZ, the interatomic potential consists of a Coulomb term to describe the long-range electrostatic interaction between ions of $Zr^{4+}$, $Y^{3+}$ and $O^{2-}$; and an empirical Born-Meyer-Buckingham (BMB) term to describe the short-range interaction between ions. The potential energy between ions $i$ and $j$, which are separated by a distance $r_{ij}$, and have ionic charges $q_i$ and $q_j$, is then given by

$$E_{ij} = \frac{q_i q_j}{r_{ij}} + A_{ij} exp\left(-\frac{r_{ij}}{\rho}\right) - \frac{C}{r_{ij}^6}, \tag{1}$$

where $A_{ij}$, $\rho$ and $C$ are potential parameters which are varied in order to reproduce experimental data. Through running simulations and comparing the results for different interatomic potential parameters in the literature [37, 41, 45, 52, 53], we have found that the interatomic potential parameters provided by Li *et al.* [53], as listed in Table 1, can better reproduce the experimental measured properties of zirconia when fitted to the lattice energy and elastic constant of zirconia. In addition, the model with this potential was found to be stable in simulations at room temperature. Therefore, this parameter set has been extensively used over the past twenty years to determine some important properties such as oxygen ionic diffusion [41, 52] and defect formation process [54] in zirconia. Our simulation results manifest that this potential model is capable of accurately simulating the dynamical behavior of YSTZ across a broad region, including the martensitic phase transformation and dislocation propagation.

Table 1. Interatomic potential parameters for the yttria-stabilized tetragonal zirconia (YSTZ) [53].

| Interaction | A(eV)   | ρ(Å)    | C(eV Å$^6$) | Mass(amu) | Charge (e) |
|-------------|---------|---------|-------------|-----------|------------|
| Zr-Zr       | 0.0     | 1.0     | 0.0         | Zr: 91.0  | Zr: +4.0   |
| Zr-Y        | 0.0     | 1.0     | 0.0         | --        | --         |
| Zr-O        | 1453.8  | 0.35    | 0.0         | --        | --         |
| Y-Y         | 0.0     | 1.0     | 0.0         | Y: 89.0   | Y: +3.0    |
| Y-O         | 826.744 | 0.35587 | 0.0         | --        | --         |
| O-O         | 22764.3 | 0.149   | 27.89       | O: 16.0   | O: -2.0    |

*2.2 Atomistic models of YSTZ*

In this section, we describe the approach that we use to generate the atomic configurations of YSTZ. The tetragonal zirconia has a space group of *P4$_2$/nmc* with two ZrO$_2$ molecules in the primitive unit cell, as shown in Fig. 1a. The lattice parameters are: $a = b = 3.64$ Å, $c = 5.27$ Å with $\alpha = \beta = \gamma = 90^0$ [55]. Then through repeating the unit cell along *a*-, *b*- and *c*-axis, a supercell of pure zirconia is generated. In comparison, the monoclinic phase of zirconia has a crystal structure with $a = 5.14$ Å, $b = 5.20$ Å, $c = 5.31$ Å, $\alpha = \gamma = 90^0$, $\beta = 99.15^0$ [56].

From the literature, it is noted that when doping with 8.0 mol% of Y$_2$O$_3$, the isothermal ionic conductivity of zirconia reaches the maximum value at low temperatures [53], and zirconia exhibits good superelasticity and shape memory properties [10, 27]. Therefore, a tetragonal zirconia single crystal stabilized by 8.0 mol% Y$_2$O$_3$ is chosen as the standard specimen in this study. To build the atomic model of YSTZ, Zr$^{4+}$ cations are randomly substituted by Y$^{3+}$ ions. In order to maintain the electrical neutrality, the substitution of Zr$^{4+}$ by Y$^{3+}$ causes the formation of oxygen vacancies, which are equal to half of Y$^{3+}$ ions. Fig. 1b presents a typical supercell of YSTZ. The system is then allowed to equilibrate to reach its minimum energy state. Simulations of uniaxial compression are carried out along some particular directions to study the orientation dependency of the failure mechanism of YSTZ. The loading process is done in a displacement-controlled manner by imposing compressive displacements to bottom-layer and top-layer atoms. The strain rate in the simulation is set to be $1 \times 10^8$/s. Constant (*N, V, T*) MD simulations are performed using the LAMMPS package [57] throughout this work. The Nose-Hoover thermostat [58] is used to maintain the temperature at a constant value of 298 K. The velocity-Verlet algorithm [59] with a time step of 1 fs is used for time integration.

## 3. Results and discussion

Unlike the ultra-high symmetric cubic YSZ, the mechanical properties of single crystalline tetragonal zirconia are anisotropic due to the distorted crystal structure. Despite the potential importance of the crystal orientation effect, only few studies have been conducted on this topic. Improving the performance of YSTZ for using in extreme conditions in various applications relies upon designing a suitable microstructure to optimize the mechanical behavior. Hence, a comprehensive understanding of the effect of crystalline orientation on mechanical properties of single crystalline YSTZ is desired. With this goal in mind, zirconia nanopillars with 11 typical crystallographic orientations, as summarized in Table 2, are generated to study the failure mechanisms of YSTZ. The simulated nanopillars specimens have the identical size of 50.0 nm × 20.0 nm × 20.0 nm and contain approximately 1.6 million atoms depending on their orientations.

Table 2. Loading direction, specimen size, and total number of atoms ($N$) in MD simulations of zirconia nanopillars.

| Sample # | Loading direction | Specimen size (nm) | Total number of atoms ($N$) |
|---|---|---|---|
| 1 | [001] | 50.0×20.0×20.0 | 1,620,000 |
| 2 | [110] | 50.0×20.0×20.0 | 1,614,354 |
| 3 | [$\bar{1}$10] | 50.0×20.0×20.0 | 1,621,080 |
| 4 | [100] | 50.0×20.0×20.0 | 1,617,980 |
| 5 | [010] | 50.0×20.0×20.0 | 1,630,200 |
| 6 | [111] | 50.0×20.0×20.0 | 1,647,562 |
| 7 | [112] | 50.0×20.0×20.0 | 1,676,918 |
| 8 | [101] | 50.0×20.0×20.0 | 1,616,945 |
| 9 | [10$\bar{1}$] | 50.0×20.0×20.0 | 1,617,220 |
| 10 | [011] | 50.0×20.0×20.0 | 1,617,110 |
| 11 | [01$\bar{1}$] | 50.0×20.0×20.0 | 1,617,000 |

From the obtained stress-strain responses, as plotted in Fig. 2, it can be seen that the YSTZ nanopillars with different orientations show rather different plastic deformation behaviors. For example, intuitively, the stress-strain curves of [101]-, [10$\bar{1}$]-, [011]-, [01$\bar{1}$]-oriented nanopillars overlap well with each other and are different from other cases. In addition, the corresponding stiffness and strength of these nanopillars are much higher than those of the other pillars, which imply distinct plastic deformation mechanisms. To unravel the mechanistic origin of the observed constitutive responses, we analyze the atomic-scale structure evolution of the material failure process. The atomistic snapshots during deformation demonstrate different failure mechanisms in different oriented nanopillars. Three types of plastic deformation mechanisms, i.e., dislocation, phase transformation and the combination of dislocation and phase transformation, are identified.

To reveal the detailed relation between failure mechanisms and constitutive responses, the stress-strain curves are categorized into three plots, as shown in Fig. 3a-c. The initial plastic deformations of [001]-, [110]-, [$\bar{1}$10]-oriented nanopillars and [101]-, [10$\bar{1}$]-, [011]-, [01$\bar{1}$]-oriented nanopillars are due to dislocation emission and martensitic ($t \rightarrow m$) phase transformation, respectively, while for the nanopillars with orientations along [100], [010], [111] and [112], dislocation together with phase transformation are observed synergize to govern the plastic deformation. Regarding the elastic behavior, all the simulated nanopillars show an initial linear elastic stage followed by a nonlinear response due to the simulation temperature of 298 K. The strength of nanopillars shows clear dependence on the failure mechanisms. Specimens dominated by dislocation emission or phase transformation exhibit the smallest and largest strength of ~15.5 GPa (Fig. 3a) and ~19.3 GPa (Fig. 3c), respectively, while when the plastic deformation mediated via a combination of dislocation and phase transformation the specimens show the moderate strength of ~16.5 GPa (Fig. 3b). Then in all cases, the stress drops abruptly bellow the elastic limit. The plastic flow stresses in Fig. 3a, c fluctuate within a certain sustainable range, while those in Fig. 3b appear to fluctuate much more severely.

To better understand the orientation effect on plastic behaviors of YSTZ, as observed in Fig. 2 and Fig. 3, the atomistic structure evolutions during compression are carefully studied in the following subsections. Three typical nanopillars with orientations along [001], [010] and [01$\bar{1}$] are selected to represent the cases with different plastic deformation mechanisms in Fig. 3a, b and c, i.e., dislocation, combination of dislocation and phase transformation, and phase transformation, respectively.

*3.1 Dislocation*

Snapshots in Fig. 4 illustrate the process of dislocation nucleation and propagation in [001]-oriented YSTZ nanopillar. From Fig. 4a, it is noted that when compressive strain increases to 0.9%, dislocation is observed to emit from one surface edge and move quickly through the pillar, which corresponds to the first abrupt stress drop in Fig. 3a (black line). When dislocation moves out of the nanopillar, surface steps are created, as denoted by blue arrows in Fig. 4a. Slip vector [60], defined as Eq. (2), is employed in Fig. 4b-d to show the dislocation propagation behavior; for clarity, atoms on perfect tetragonal lattice are not shown in the plots.

$$S^\alpha = -\frac{1}{n_s}\Sigma_{\beta \neq \alpha}^n (x^{\alpha\beta} - X^{\alpha\beta}) \tag{2}$$

In the above expression, *n* is the number of nearest neighbors to atom *α*, $n_s$ is the number of slipped neighbors, and $x^{\alpha\beta}$ and $X^{\alpha\beta}$ are the vector differences of atoms *α* and *β* current and reference positions, respectively. The reference configuration is the arrangement of atomic position associated with zero mechanical stress.

It can be seen in Fig. 4b that the initial dislocation appears in the form of partial dislocation loops with stacking fault left after migration. The dislocation nucleates at the surface edge of the pillar and then traverse through the whole nanopillar along the [111] direction with $\frac{1}{2}\langle 111 \rangle$ Burgers vector (3.68 Å). Typically ⟨111⟩ dislocations form at corners, which allow them to be short and along with a mixed edge-screw character. Closer inspection of the slip plane atoms (Fig. 4b-d) reveals that the dislocation lines and slip planes are curvy rather than straight and flat, which implies the character of mixed edge and screw. In our simulations the dislocation migration is very rapid, which makes it hard to identify individual fractional dislocation events. With further compression, the stress in Fig. 3a increases again with a small magnitude to about 9.0 GPa at strain of 1.0%, which corresponds to the second Burgers vector that dislocation attempts to migrate on. For a specific dislocation event, the energy arrogation before emission will cause a local increase of stress, and then the energy release after propagation will lead to the next local drop of stress. As a result, the plastic stress-strain response of [001]-oriented nanopillar shows a serrated and locally periodic characteristic (Fig. 3a), which is not observed in the cases of phase transformation (Fig. 3c). With the migration of the primary dislocation, however, a new dislocation is noticed to emit from another edge of the nanopillar, as shown in Fig. 4c, and then propagate almost perpendicular to the loading direction. With the increase of strain, more and more dislocations are triggered and emit from the internal sites, as shown in Fig. 4d. These dislocations glide along different slip planes, and meet other newly formed dislocations. Each dislocation slip event creates a step on the surface of the nanopillar, which leads the nanopillar to be much rougher. Less stress required to initiate subsequent bursts, which are due to the fact that the surface roughening of nanopillar has lowered the potential energy barrier for dislocation creation at the surface, which is also responsible for the decrease of the magnitude of local periodic peak during the plastic flow. It is worth noticing that the primary slip planes are along the ⟨111⟩ direction, and no evidence of phase transformation is observed. Similar dislocation glide behavior is observed in the [110]- and [$\bar{1}$10]-oriented YSTZ nanopillars, which is in good agreement with a recent experimental observation showing that a crystal slip is preferentially triggered when the orientations of pillars are near [110] direction [49].

*3.2 Phase Transformation*

For the case of phase transformation in nanopillars, a sudden stress drop is observed at strain ~1.1% (Fig. 3c). Analysis of atomic trajectories reveals that a deformation behavior different than dislocation is induced. Fig. 5a shows side view of a deformed atomic configuration of [01$\bar{1}$]-oriented YSTZ nanopillar at strain of ε = 2.6%. It should be noted that the atoms residing in a tilted band, as shown in Fig. 5a, undergo considerable crystalline structure re-arrangement, indicating the initiation of $t \rightarrow m$ martensitic phase transformation. At this stage, the tetragonal phase adjacent to the region of the tilted band starts to transform into the monoclinic phase. This transformed phase can be verified by a careful inspection of the crystal structure of the transformed region, as presented in Fig. 5c. The transformed monoclinic phase can be clearly seen, and its structure is different than the original tetragonal phase (Fig. 5b). One of the lattice variables, the angle showing in Fig. 5c, is measured to be $99.5^0$, which is in a good agreement with the theoretical prediction of the angle of monoclinic phase ($99.15^0$) [56]. To validate the transformed phase, radial distribution functions (RDF) g(r) (where r is the distance between two cations) of cation pairs (Zr-Zr) and (Zr-O) of the perfect tetragonal and monoclinic phases are plotted and carefully compared with that of the transformed phase, as it can be seen in Fig. 6a and Fig. 6b. The site of the nearest Zr-Zr neighbor of the transformed phase, namely the location of first peak of RDF curve, is at 3.25 Å (red line), which is much closer to that of the perfect monoclinic phase (3.45 Å, black line) than that of the tetragonal phase (3.65 Å, blue line). It is worth noting that the transformed phase is elastically compressed along the applied loading direction, therefore, it is reasonable to imply that the transformed phase is the compressed monoclinic zirconia. Additionally, from Fig. 6b it can be noted that the first and second nearest neighbors of Zr-O pair of the transformed phase (red line) is in between those of the tetragonal (blue line) and the monoclinic (black line) phases. The distance between first and second nearest neighbors of Zr-O pair in transformed phase becomes shorter, which is much similar to the short distance in monoclinic phase rather than the long space in tetragonal phase, as shown in Fig. 6b. Moreover, in the transformed phase, the atom density g(r) of the first peak is smaller than that of the second peak, and this pattern is similar to that of the monoclinic phase, but opposite of the tetragonal phase. This observed phenomenon is the direct evidence of $t \rightarrow m$ phase transformation that is reported experimentally [10, 27].

According to the location of the nearest Zr-Zr neighbor of tetragonal zirconia (3.65 Å) and transformed phase (3.25 Å), as shown in Fig. 6a, a cutoff distance of 3.5 Å, which is in between of 3.65 Å and 3.25 Å, is chosen to calculate the coordination number (CN) of Zr atoms. Fig. 7a-d present the atomistic snapshots of the deformed $[01\bar{1}]$-oriented single crystalline YSTZ nanopillar at various stages of compression. In these pictures, the coordinate number (CN) is used to illustrate the process of phase transformation. Different variants of the Y-$ZrO_2$ structure can be discerned. Atoms on perfect tetragonal lattice are shown in blue color (CN = 0), while the transformed monoclinic phase is in yellow-green color (CN = 6). To furthermore identify the transformed phase, in Fig. 8, the coordinate number of Zr atoms within the same cutoff of 3.5 Å in monoclinic phase is counted to be 6 (i.e., 3.36 Å (1), 3.45 Å (2), 3.48 Å (3), as highlighted with box in Fig. 8), which is again in consistent with that of the transformed phase. The tilted band region of transformed phase is observed to initiate from the center of the model (Fig. 7a) and then expand oppositely from both sides of the tilted band to the top and bottom surfaces of the nanopillar (Fig. 7b). It is worth mentioning that the transformed phase band is not exactly perpendicular to the compressive loading, and rotates an angle of about $9.2^0$ to the (100) top/bottom surface (Fig. 5a and Fig. 7a), which is consistent with the observation from experiment [2] and phase field simulation [61]. With the strain increases to 3.6%, a clear surface corrugation is observed due to the expansion of volume during phase transformation, as denoted by black arrows in Fig. 7c-d. This observation again supports the experiment results that the $t \rightarrow m$ phase transformation is accompanied by a volume expansion [21]. Indeed, such surface relief was also observed in the AFM experiment of indentation on stabilized zirconia [21]. On the other hand, such volume expansion is also responsible for the stress increase after plastic flow, as can be seen in Fig. 3c. Furthermore, a new region of phase transformation is formed at the bottom of the nanopillar, which expands with the increase of loading. A local sudden drop of stress is noticed in the stress-strain curve, as denoted by a dotted ellipsoid in Fig. 3c. It is believed that this transformed monoclinic phase region is accounted for the high local stress, while the surface corrugation induced by the transformed phase is responsible for the release of strain energy. As the strain reaches ~ 4.8%, some of the internal area of transformed phase cannot sustain the increasing compressive loading and undergoes an additional plastic deformation, as indicated by an ellipsoid in Fig. 7d. Such failure within transformed phase is commonly observed, such as in silicon [51].

It is finally worth mentioning that the stress-strain responses of the nanopillars undergoing a phase transformation, i.e., [101]-, [10$\bar{1}$]-, [011]-, [01$\bar{1}$]-oriented nanopillars, are very similar to each other. This implies that the material is isotropic along these four crystallographic orientations, which is consistent to the experiment explored crystalline structure of tetragonal zirconia [25]. In contrast, stress is more sensitive to the dislocation emission in both elastic and plastic flow regions (Fig. 3a) due to the series of local energy release.

*3.3 Combination of Dislocation and Phase Transformation*

When applying load along [100], [010], [111] or [112] directions, the plastic deformation of nanopillar is observed to be a combination of dislocation migration and $t \rightarrow m$ phase transformation. Consider the [010]-oriented nanopillar for instance, Fig. 9 presents the atomistic configurations of its deformed structure with the appearance of dislocation and phase transformation. The transformed phase shows highly crystal order (Fig. 9c) rather than amorphous. Rdf of Zr-Zr pair (Fig. 9d) demonstrates that the first peak of transformed phase, namely the nearest neighbor, locates at ~3.2 Å, which is much closer to that of the monoclinic phase (3.45 Å) than that of tetragonal phase (3.65 Å). Therefore, it is reasonable to conclude that the transformed phase is a polymorphous of the compressed monoclinic Y-ZrO$_2$.

The process of dislocation migration and phase transformation is elaborated by coordinate number (CN), as presented in Fig. 10. It can be seen from Fig. 10a that when the compressive strain increases to 1.0%, two dislocations have emitted from points *A* and *B*, and propagated along the (011) slip planes. This behavior is reflected as the sudden stress drop in the stress-strain curve (Fig. 3b). With continued loading, the primary dislocation along *AC* will attempt to overcome the stacking fault energy on slip plane to migrate another Burgers vector, by which leading partial and trailing partial are observed in Fig. 10b-d. However, subsequently, the motion of dislocation, which emits from point *B*, is obstructed by formation of the transformed phase, as denoted by green color in Fig. 10b-d. The obstruction by phase transformation leads to a severe challenge for dislocation to move, then results in the local stress jump at strain ~2.0% in the plastic flow stage (see Fig. 3b). The transformed phase region expands with the continued loading. With the accumulation of internal energy, dislocation finally conquers the barrier of transformed phase region and moves out of the specimen at point *D* (Fig. 10c). Due to the disturbance of phase transformation, only partial dislocation is induced along *BD* direction. Spontaneously, a new dislocation nucleates from point *E* and propagates parallel to dislocation *AC* until the emission of another dislocation from point *F*, and then it moves along [01$\bar{1}$] direction. Eventually, these two dislocation lines meet at point *G*, which blocks the subsequent motion of dislocation (Fig. 10d). It is worth noting that with the increase of loading, the failure of the transformed phase is observed, as can be identified as the yellow atoms within the region of transformed phase in Fig. 10d, which is similar to the case of pure phase transformation in Fig. 7d. Surface defects (steps) are left at points of *A*, *B*, *C*, *D*, *E* and *F* due to the migration of dislocation, which is an indication of energy release. Moreover, due to the competition of dislocation propagation and phase transformation, the stress in plastic region show severe fluctuation (Fig. 3b) than that of pure dislocation (Fig. 3a) or pure phase transformation (Fig. 3c) cases. In addition, it is noted from Fig. 10b-d that the Zr atom within the transformed phase has a coordinate number (CN) of 6 (green colored atoms), which is in a good agreement with that of the monoclinic phase. Eventually, it is worth to emphasis that for the nanopillars discussed in this section although both dislocation and phase transformation are observed, dislocation migration happens first and therefore serve as the leading failure mechanism, which is responsible for the significant strength drop in contrast to the pure phase transformation cases (c.f. Fig. 3b, Fig. 3c), but the strength increases slightly in contrast to the pure dislocation migration cases (c.f. Fig. 3a, Fig. 3b).

## 4. Conclusions

In this study, atomistic simulations were utilized to investigate the plastic deformation mechanisms of single crystalline yttria-stabilized tetragonal zirconia (YSTZ) nanopillars under uniaxial compression. Given the relatively low symmetry characteristic of tetragonal zirconia, orientation effect on mechanical properties of YSTZ nanopillars is potentially important, and it has been comprehensively studied in this work. Three typical plastic deformation behaviors, i.e., dislocation propagation, phase transformation and a combination of dislocation and phase transformation, were identified when applying compressive loading along different directions (Table 3). To the best of our knowledge, this is the first time that the experimental observed $t \rightarrow m$ martensitic phase transformation is reproduced by MD simulations.

Table 3. Summary of zirconia nanopillars with different crystallographic orientations including loading direction, specimen size, total number of atoms (*N*), simulated compressive strength, and the observed dominated plastic deformation mechanisms.

| Specimen # | Loading direction | Strength (GPa) | Dominated Deformation Mechanism |
| --- | --- | --- | --- |
| 1 | [001] | 15.0GPa | Dislocation |
| 2 | [110] | 15.5GPa | Dislocation |
| 3 | [$\bar{1}$10] | 15.6GPa | Dislocation |
| 4 | [100] | 16.2GPa | Dislocation + Phase transformation |
| 5 | [010] | 16.8GPa | Dislocation + Phase transformation |
| 6 | [111] | 16.7GPa | Dislocation + Phase transformation |
| 7 | [112] | 16.8GPa | Dislocation + Phase transformation |
| 8 | [101] | 19.3GPa | Phase transformation |
| 9 | [10$\bar{1}$] | 19.3GPa | Phase transformation |
| 10 | [011] | 19.3GPa | Phase transformation |
| 11 | [01$\bar{1}$] | 19.3GPa | Phase transformation |

The nanoscale plastic deformation of YSTZ showed a strong dependence on the crystallographic orientations of nanopillars. According to our simulation results, as summarized in Table 3, the [001]-, [110]- and [$\bar{1}$10]-oriented nanopillars failed by dislocation emission and propagation. {111} plane was observed as the primary slip plane. For the orientations of [101], [10$\bar{1}$], [011] or [01$\bar{1}$], tetragonal (*t*) to monoclinic (*m*) phase transformation was identified as the dominant deformation mechanism in nanopillars. The transformed phase band tilted to the top/bottom surfaces of the nanopillars with an angle of ~9.2$^0$, which is in good agreement with experimental observations [2]. When the applied loading was along [100]-, [010]-, [111]- and [112], a combination of dislocation migration and $t \rightarrow m$ phase transformation was explored. The competition between dislocation motion and phase transformation led to a severe fluctuation of plastic stress flow. The strength of zirconia nanopillar exhibited a sensitive behavior on the plastic deformation mechanisms, i.e., dislocation- and phase transformation-dominated deformations led to the lowest and highest strength, ~15.5 GPa vs. ~19.3 GPa, respectively; while deformation mediated by the combination of dislocation and phase transformation resulted in the moderate strength, i.e. ~16.5 GPa.

Overall, our simulation results provided an atomistic viewpoint of the plastic deformation mechanisms of YSTZ nanopillars under compressive loading. The new insights about crystallographic orientation effect gained in this study have improved our understanding of the complex deformation mechanisms of zirconia nanopillars, and this potentially enables a reliable prediction and provides guidelines in the microstructural design of zirconia and similar materials undergoing martensitic phase transformation.

**Acknowledgment**

The authors are grateful for computer time allocation provided by the Extreme Science and Engineering Discovery Environment (XSEDE).

**References**

[1] M. Borik, V. Bublik, M. Vishnyakova, E. Lomonova, V. Myzina, N.Y. Tabachkova, A. Timofeev, Structure and phase composition studies of partially stabilized zirconia, Journal of Surface Investigation. X-ray, Synchrotron and Neutron Techniques 5(1) (2011) 166-171.
[2] R.H. Hannink, P.M. Kelly, B.C. Muddle, Transformation toughening in zirconia-containing ceramics, Journal of the American Ceramic Society 83(3) (2000) 461-487.
[3] E.J. Walter, S.P. Lewis, A.M. Rappe, First principles study of carbon monoxide adsorption on zirconia-supported copper, Surface science 495(1) (2001) 44-50.
[4] K. Tanabe, T. Yamaguchi, Acid-base bifunctional catalysis by $ZrO_2$ and its mixed oxides, Catalysis Today 20(2) (1994) 185-197.
[5] P. Charpentier, P. Fragnaud, D. Schleich, E. Gehain, Preparation of thin film SOFCs working at reduced temperature, Solid State Ionics 135(1) (2000) 373-380.

[6] D. Clarke, C. Levi, Materials design for the next generation thermal barrier coatings, Annual Review of Materials Research 33(1) (2003) 383-417.
[7] A. Meldrum, L. Boatner, R. Ewing, Nanocrystalline zirconia can be amorphized by ion irradiation, Physical review letters 88(2) (2001) 025503.
[8] V. Fiorentini, G. Gulleri, Theoretical evaluation of zirconia and hafnia as gate oxides for Si microelectronics, Physical review letters 89(26) (2002) 266101.
[9] A. Lai, Z. Du, C.L. Gan, C.A. Schuh, Shape memory and superelastic ceramics at small scales, Science 341(6153) (2013) 1505-1508.
[10] Z. Du, X.M. Zeng, Q. Liu, A. Lai, S. Amini, A. Miserez, C.A. Schuh, C.L. Gan, Size effects and shape memory properties in ZrO 2 ceramic micro-and nano-pillars, Scripta Materialia 101 (2015) 40-43.
[11] N.R. Silva, I. Sailer, Y. Zhang, P.G. Coelho, P.C. Guess, A. Zembic, R.J. Kohal, Performance of zirconia for dental healthcare, Materials 3(2) (2010) 863-896.
[12] T. Gupta, J. Bechtold, R. Kuznicki, L. Cadoff, B. Rossing, Stabilization of tetragonal phase in polycrystalline zirconia, Journal of Materials Science 12(12) (1977) 2421-2426.
[13] R. Garvie, Stabilization of the tetragonal structure in zirconia microcrystals, The Journal of Physical Chemistry 82(2) (1978) 218-224.
[14] D.A. Ward, E.I. Ko, Synthesis and structural transformation of zirconia aerogels, Chemistry of materials 5(7) (1993) 956-969.
[15] T. Höche, M. Deckwerth, C. Rüssel, Partial Stabilization of Tetragonal Zirconia in Oxynitride Glass-Ceramics, Journal of the American Ceramic Society 81(8) (1998) 2029-2036.
[16] S. Tsunekawa, S. Ito, Y. Kawazoe, J.-T. Wang, Critical size of the phase transition from cubic to tetragonal in pure zirconia nanoparticles, Nano Letters 3(7) (2003) 871-875.
[17] R. Garvie, R. Hannink, R. Pascoe, Ceramic steel?, (1975).
[18] C.M. Wayman, Introduction to the crystallography of martensitic transformations, Macmillan1964.
[19] N. Ross, A. Crocker, A generalized theory of martensite crystallography and its application to transformations in steels, Acta Metallurgica 18(4) (1970) 405-418.
[20] A. Acton, M. Bevis, A generalised martensite crystallography theory, Materials Science and Engineering 5(1) (1969) 19-29.
[21] S. Deville, G. Guénin, J. Chevalier, Martensitic transformation in zirconia: part II. Martensite growth, Acta materialia 52(19) (2004) 5709-5721.
[22] T. Sato, M. Shimada, Transformation of Yttria-Doped Tetragonal ZrO2 Polycrystals by Annealing in Water, Journal of the American Ceramic Society 68(6) (1985) 356-356.
[23] H. El Kadiri, Z. Utegulov, M. Khafizov, M.A. Zaeem, M. Mamivand, A. Oppedal, K. Enakoutsa, M. Cherkaoui, R. Graham, A. Arockiasamy, Transformations and cracks in zirconia films leading to breakaway oxidation of Zircaloy, Acta Materialia 61(11) (2013) 3923-3935.
[24] K. Edalati, S. Toh, Y. Ikoma, Z. Horita, Plastic deformation and allotropic phase transformations in zirconia ceramics during high-pressure torsion, Scripta Materialia 65(11) (2011) 974-977.
[25] K. Matsui, H. Yoshida, Y. Ikuhara, Phase-transformation and grain-growth kinetics in yttria-stabilized tetragonal zirconia polycrystal doped with a small amount of alumina, Journal of the European Ceramic Society 30(7) (2010) 1679-1690.

[26] M. Trunec, Z. Chlup, Higher fracture toughness of tetragonal zirconia ceramics through nanocrystalline structure, Scripta Materialia 61(1) (2009) 56-59.
[27] J. San Juan, M. Nó, Superelasticity and shape memory at nano-scale: Size effects on the martensitic transformation, Journal of Alloys and Compounds 577 (2013) S25-S29.
[28] L. Huanga, X.-L. Yuana, S.-X. Cuia, D.-Q. Weia, The compression behaviors of zirconium from the first-principle calculations, (2013).
[29] G. Stapper, M. Bernasconi, N. Nicoloso, M. Parrinello, Ab initio study of structural and electronic properties of yttria-stabilized cubic zirconia, Physical Review B 59(2) (1999) 797.
[30] M. Mamivand, M.A. Zaeem, H. El Kadiri, Phase field modeling of stress-induced tetragonal-to-monoclinic transformation in zirconia and its effect on transformation toughening, Acta Materialia 64 (2014) 208-219.
[31] M. Mamivand, M.A. Zaeem, H. El Kadiri, Shape memory effect and pseudoelasticity behavior in tetragonal zirconia polycrystals: A phase field study, International Journal of Plasticity 60 (2014) 71-86.
[32] M. Mamivand, M.A. Zaeem, H. El Kadiri, Effect of variant strain accommodation on the three-dimensional microstructure formation during martensitic transformation: Application to zirconia, Acta Materialia 87 (2015) 45-55.
[33] M. Mamivand, M.A. Zaeem, H. El Kadiri, A review on phase field modeling of martensitic phase transformation, Computational Materials Science 77 (2013) 304-311.
[34] H.B. Lee, F.B. Prinz, W. Cai, Atomistic simulations of grain boundary segregation in nanocrystalline yttria-stabilized zirconia and gadolinia-doped ceria solid oxide electrolytes, Acta Materialia 61(10) (2013) 3872-3887.
[35] C. Retamal, M. Lagos, B.M. Moshtaghioun, F.L. Cumbrera, A. Domínguez-Rodríguez, D. Gómez-García, A new approach to the grain-size dependent transition of stress exponents in yttria tetragonal zirconia polycrystals. The theoretical limit for superplasticity in ceramics, Ceramics International (2015).
[36] S. Fabris, A.T. Paxton, M.W. Finnis, A stabilization mechanism of zirconia based on oxygen vacancies only, Acta Materialia 50(20) (2002) 5171-5178.
[37] P.K. Schelling, S.R. Phillpot, Mechanism of Thermal Transport in Zirconia and Yttria-Stabilized Zirconia by Molecular-Dynamics Simulation, Journal of the American Ceramic Society 84(12) (2001) 2997-3007.
[38] T. Arima, K. Fukuyo, K. Idemitsu, Y. Inagaki, Molecular dynamics simulation of yttria-stabilized zirconia between 300 and 2000 K, Journal of Molecular Liquids 113(1) (2004) 67-73.
[39] H. Hayashi, T. Saitou, N. Maruyama, H. Inaba, K. Kawamura, M. Mori, Thermal expansion coefficient of yttria stabilized zirconia for various yttria contents, Solid State Ionics 176(5) (2005) 613-619.
[40] T. Tojo, H. Kawaji, T. Atake, Molecular dynamics study on lattice vibration and heat capacity of yttria-stabilized zirconia, Solid State Ionics 118(3) (1999) 349-353.
[41] H. Brinkman, W. Briels, H. Verweij, Molecular dynamics simulations of yttria-stabilized zirconia, Chemical physics letters 247(4) (1995) 386-390.
[42] R. Krishnamurthy, Y.G. Yoon, D. Srolovitz, R. Car, Oxygen Diffusion in Yttria-Stabilized Zirconia: A New Simulation Model, Journal of the American Ceramic Society 87(10) (2004) 1821-1830.
[43] Y. Yamamura, S. Kawasaki, H. Sakai, Molecular dynamics analysis of ionic conduction mechanism in yttria-stabilized zirconia, Solid State Ionics 126(1) (1999) 181-189.

[44] W. Araki, Y. Arai, Molecular dynamics study on oxygen diffusion in yttria-stabilized zirconia subjected to uniaxial stress in terms of yttria concentration and stress direction, Solid State Ionics 181(33) (2010) 1534-1541.
[45] P.K. Schelling, S.R. Phillpot, D. Wolf, Mechanism of the Cubic-to-Tetragonal Phase Transition in Zirconia and Yttria-Stabilized Zirconia by Molecular-Dynamics Simulation, Journal of the American Ceramic Society 84(7) (2001) 1609-1619.
[46] U. Messerschmidt, D. Baither, B. Baufeld, M. Bartsch, Plastic deformation of zirconia single crystals: a review, Materials Science and Engineering: A 233(1) (1997) 61-74.
[47] S. Azad, O.A. Marina, C.M. Wang, L. Saraf, V. Shutthanandan, D.E. McCready, A. El-Azab, J.E. Jaffe, M.H. Engelhard, C.H. Peden, Nanoscale effects on ion conductance of layer-by-layer structures of gadolinia-doped ceria and zirconia, Applied Physics Letters 86(13) (2005) 131906.
[48] H. Yoshida, K. Yokoyama, N. Shibata, Y. Ikuhara, T. Sakuma, High-temperature grain boundary sliding behavior and grain boundary energy in cubic zirconia bicrystals, Acta materialia 52(8) (2004) 2349-2357.
[49] X.M. Zeng, A. Lai, C.L. Gan, C.A. Schuh, Crystal orientation dependence of the stress-induced martensitic transformation in zirconia-based shape memory ceramics, Acta Materialia 116 (2016) 124-135.
[50] N. Zhang, Y. Chen, Nanoscale plastic deformation mechanism in single crystal aragonite, Journal of Materials Science 48(2) (2013) 785-796.
[51] N. Zhang, Q. Deng, Y. Hong, L. Xiong, S. Li, M. Strasberg, W. Yin, Y. Zou, C.R. Taylor, G. Sawyer, Deformation mechanisms in silicon nanoparticles, Journal of Applied Physics 109(6) (2011) 063534.
[52] M. Kilo, M. Taylor, C. Argirusis, G. Borchardt, R. Jackson, O. Schulz, M. Martin, M. Weller, Modeling of cation diffusion in oxygen ion conductors using molecular dynamics, Solid State Ionics 175(1) (2004) 823-827.
[53] X. Li, B. Hafskjold, Molecular dynamics simulations of yttrium-stabilized zirconia, Journal of Physics: Condensed Matter 7(7) (1995) 1255.
[54] X. Wang, X. Liu, A. Javed, C. Zhu, G. Liang, Phase transition behavior of yttria-stabilized zirconia from tetragonal to monoclinic in the lanthanum zirconate/yttria-stabilized zirconia coupled-system using molecular dynamics simulation, Journal of Molecular Liquids 207 (2015) 309-314.
[55] G. Teufer, The crystal structure of tetragonal ZrO2, Acta Crystallographica 15(11) (1962) 1187-1187.
[56] J.t. McCullough, K. Trueblood, The crystal structure of baddeleyite (monoclinic ZrO2), Acta Crystallographica 12(7) (1959) 507-511.
[57] S. Plimpton, Fast parallel algorithms for short-range molecular dynamics, Journal of computational physics 117(1) (1995) 1-19.
[58] W.G. Hoover, Canonical dynamics: equilibrium phase-space distributions, Physical Review A 31(3) (1985) 1695.
[59] L. Verlet, Computer" experiments" on classical fluids. I. Thermodynamical properties of Lennard-Jones molecules, Physical review 159(1) (1967) 98.
[60] J. Zimmerman, C. Kelchner, P. Klein, J. Hamilton, S. Foiles, Surface step effects on nanoindentation, Physical Review Letters 87(16) (2001) 165507.

[61] M. Mamivand, M.A. Zaeem, H. El Kadiri, L.-Q. Chen, Phase field modeling of the tetragonal-to-monoclinic phase transformation in zirconia, Acta Materialia 61(14) (2013) 5223-5235.

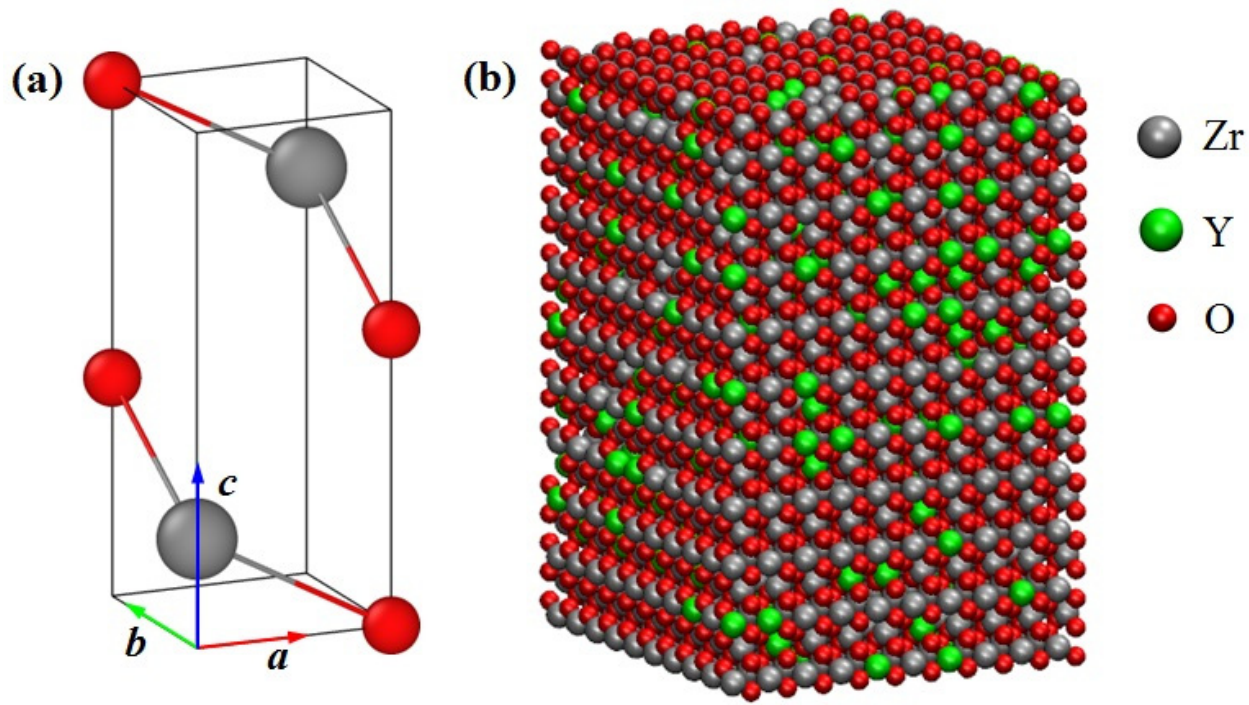

Fig. 1. (a) The primitive unit cell of tetragonal $ZrO_2$; (b) A supercell of yttria-stabilized tetragonal zirconia (YSTZ).

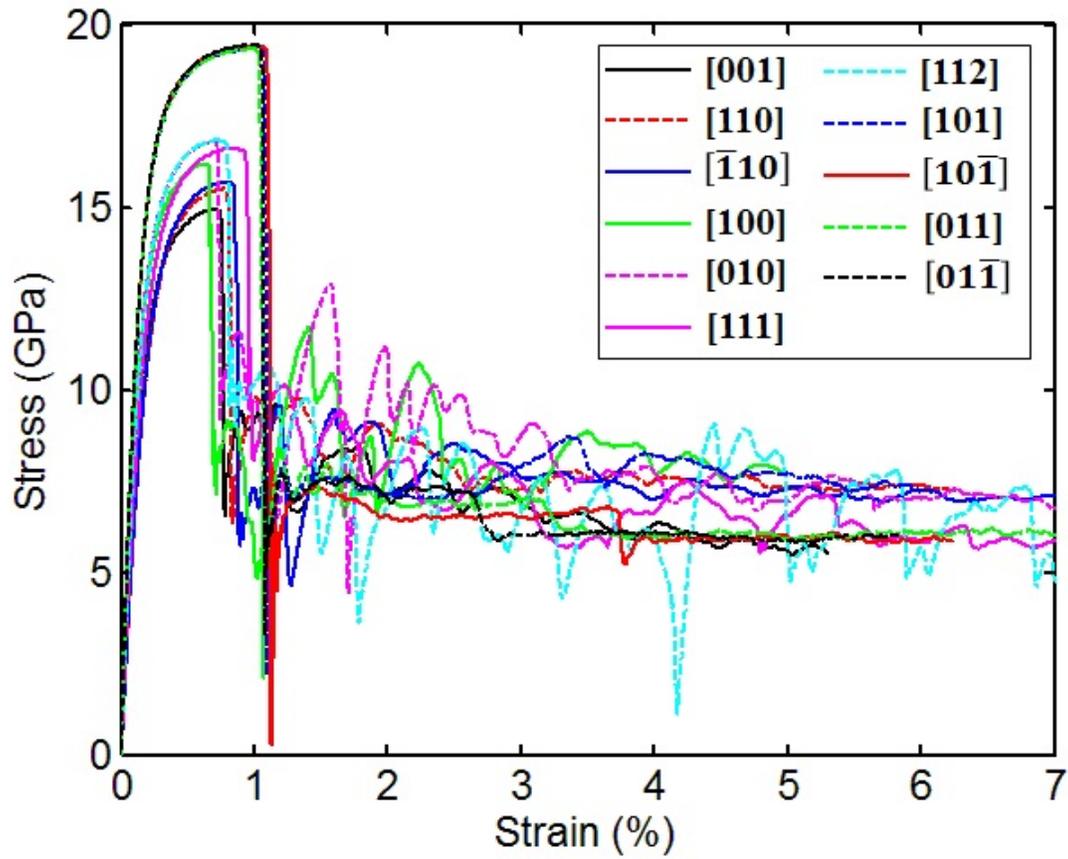

Fig. 2. Stress-strain relations of 11 YSTZ nanopillars with different orientations under uniaxial compressive loading.

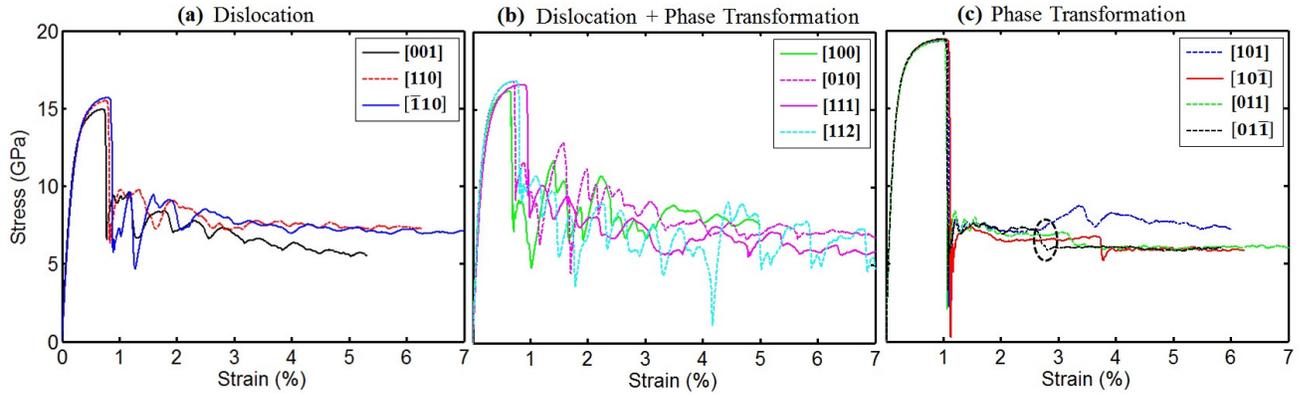

Fig. 3. Stress-strain curves of YSTZ nanopillars under compressive loading along different orientations: (a) [001], [110] and [$\bar{1}$10]; (b) [010], [100], [112] and [111]; (c) [101], [011], [10$\bar{1}$] and [01$\bar{1}$]. The stress-strain curves are categorized according to the observed deformation mechanisms.

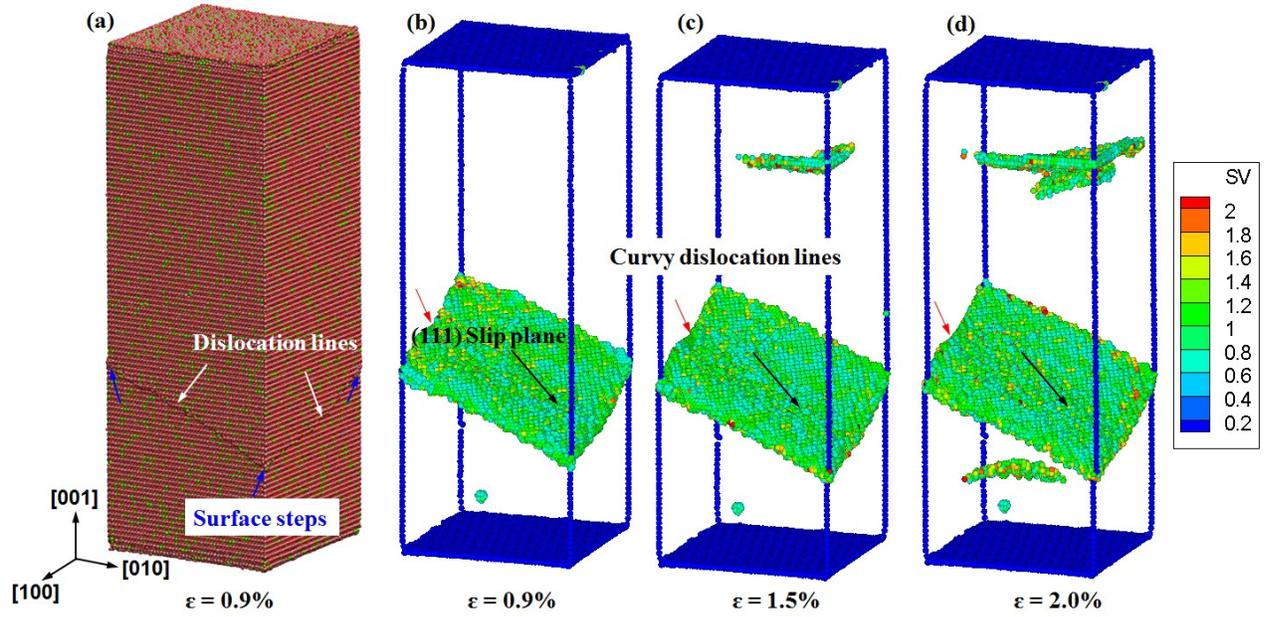

Fig. 4. 3D atomistic configurations of [001]-oriented zirconia nanopillar regarding dislocation propagation process at strains of (a-b): ε = 0.9%, (c) ε = 1.5% and (d) ε = 2.0%. Slip vector (SV) is adopted in (b-d) to track dislocation migration process. The colorbar represents the value of slip vector for each atom.

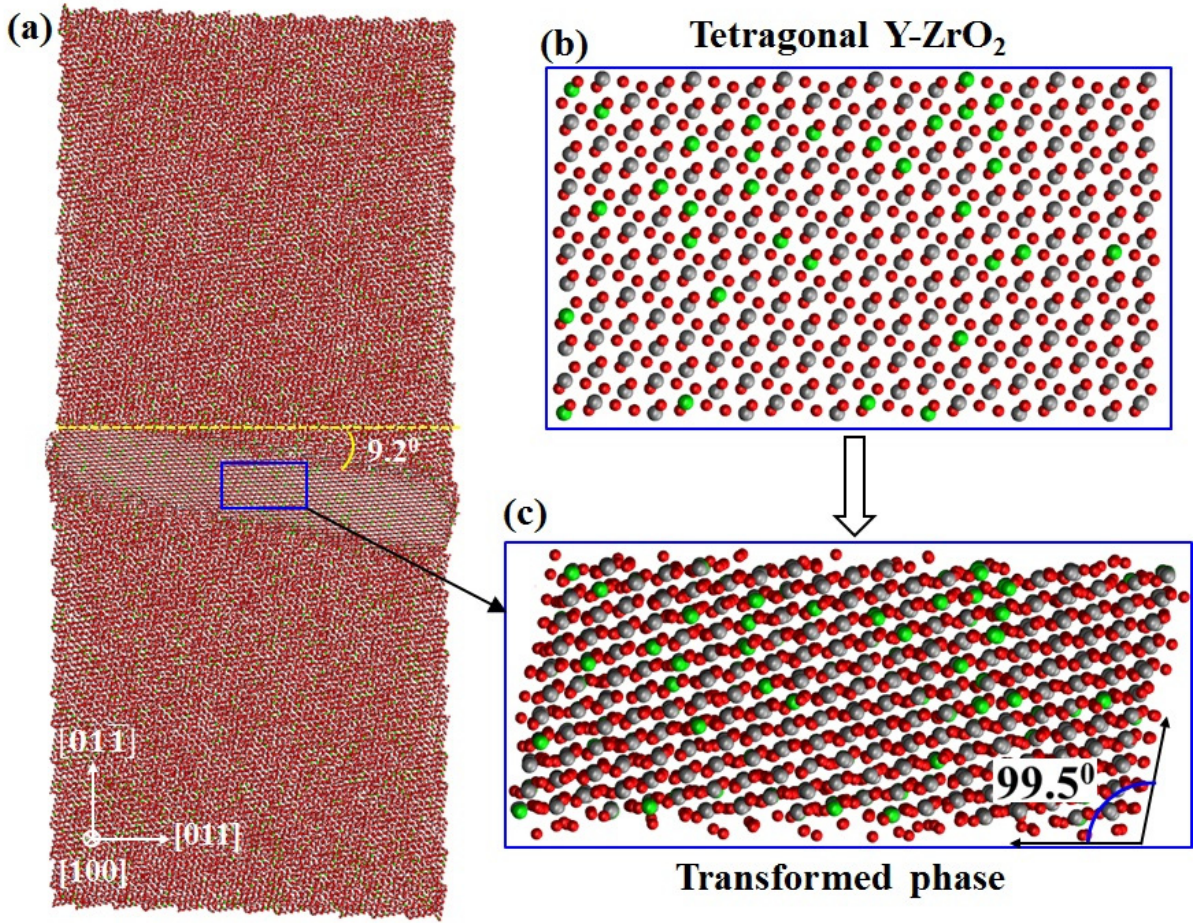

Fig. 5. Phase transformation is observed in the middle tilted band of a single crystalline YSTZ nanopillar with uniaxial compression along $[01\bar{1}]$ direction. (a) Side view of the deformed nanopillar; (b) The original structure of tetragonal Y-ZrO$_2$; (c) The structure of transformed monoclinic phase.

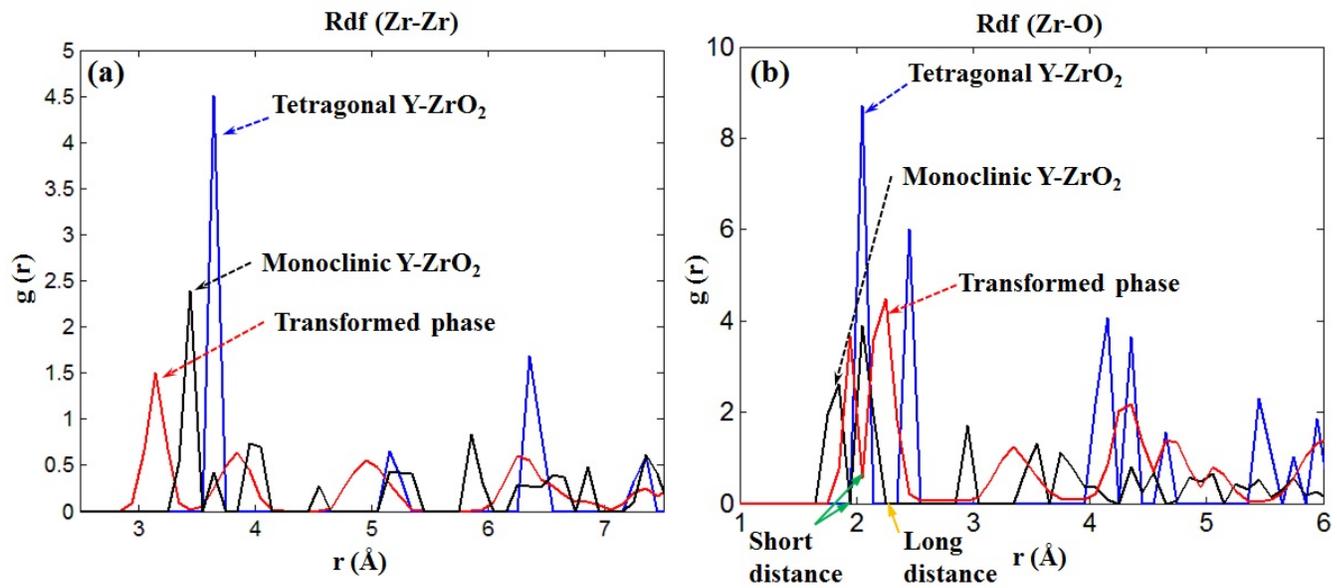

Fig. 6. Comparisons of radial distribution function (RDF) (a) Zr-Zr and (b) Zr-O of tetragonal, monoclinic and transformed phase.

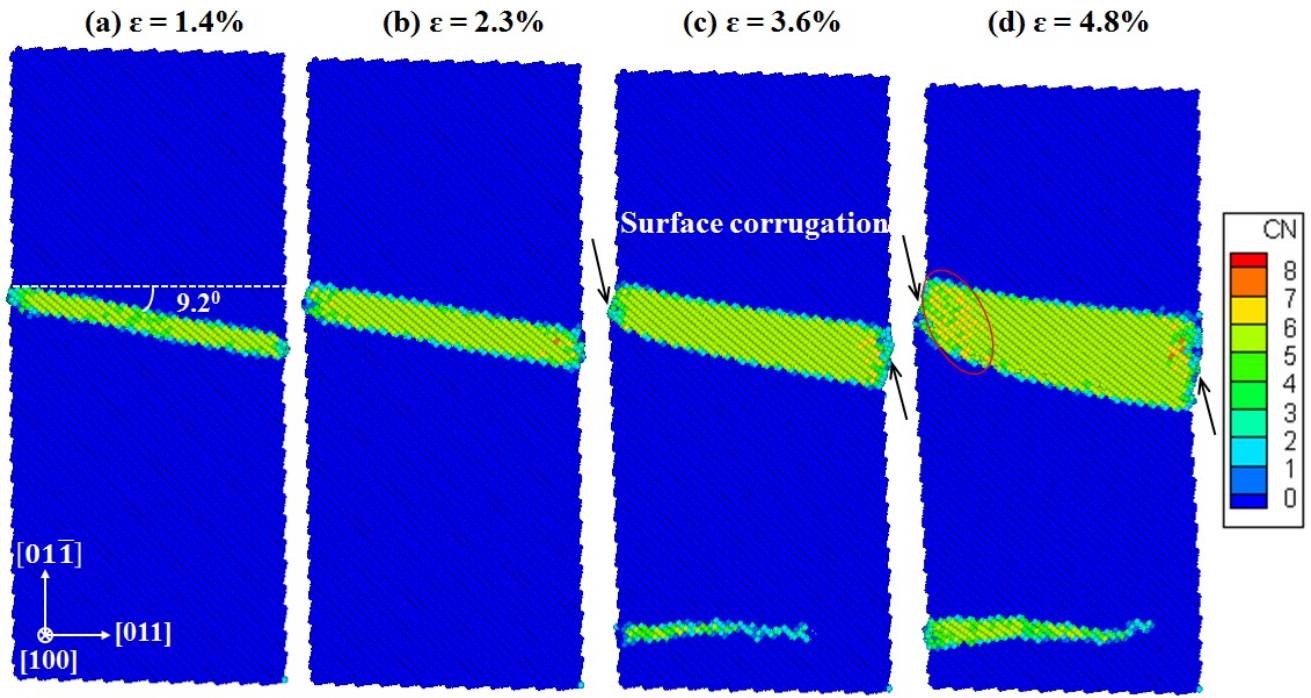

Fig. 7. The evolution of phase transformation in [01$\bar{1}$]-oriented YSTZ nanopillar at uniaxial compressive strains of (a) ε = 1.4%; (b) ε = 2.3%; (c) ε = 3.6% and (d) ε = 4.8%. Atoms are colored by coordination number (CN).

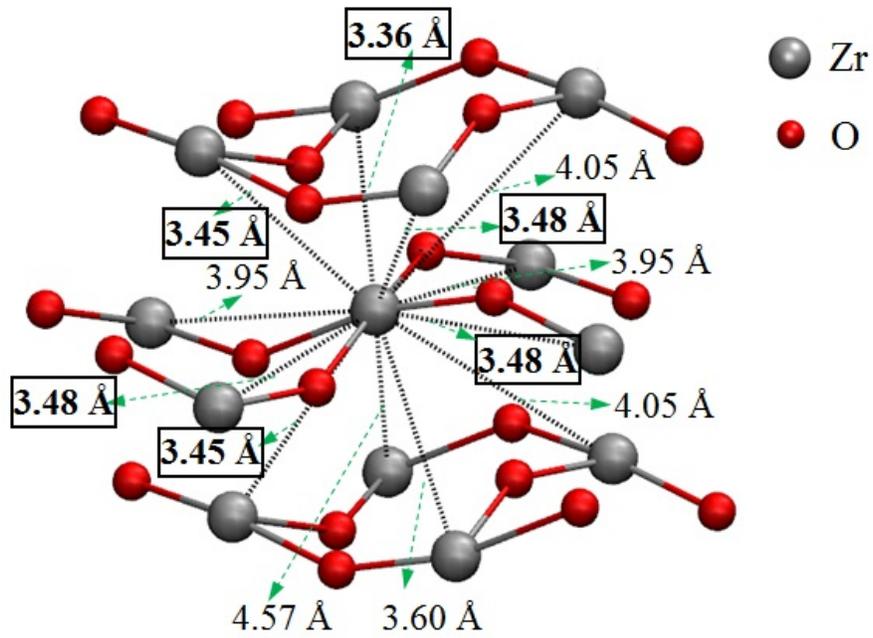

Fig. 8. Structure of monoclinic $ZrO_2$ phase. The distances between Zr-Zr neighbors are measured.

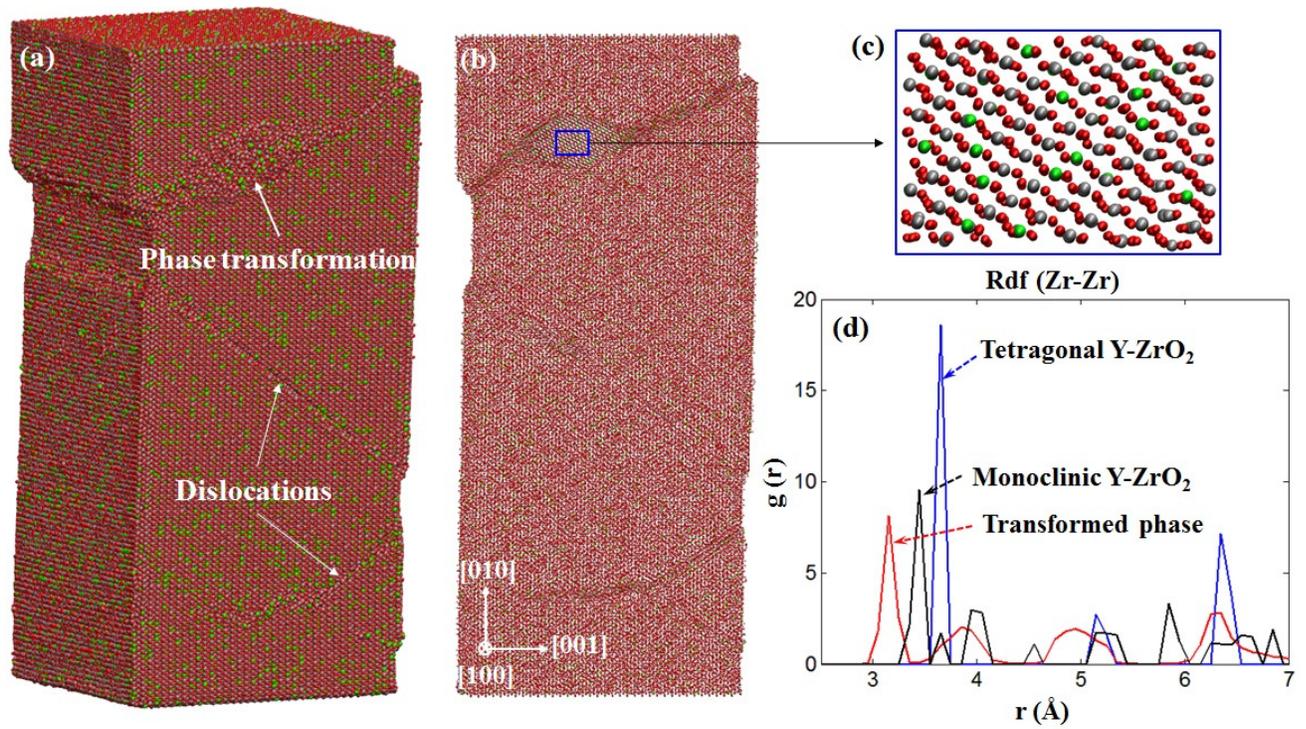

Fig. 9. (a) 3D and (b) side view of the atomistic configuration of deformed [010]-oriented YSTZ nanopillar. (c) Crystal structure of the transformed phase. (d) A comparison of Rdf (Zr-Zr) in the phases of tetragonal, monoclinic and the transformed phase of Y-$ZrO_2$.

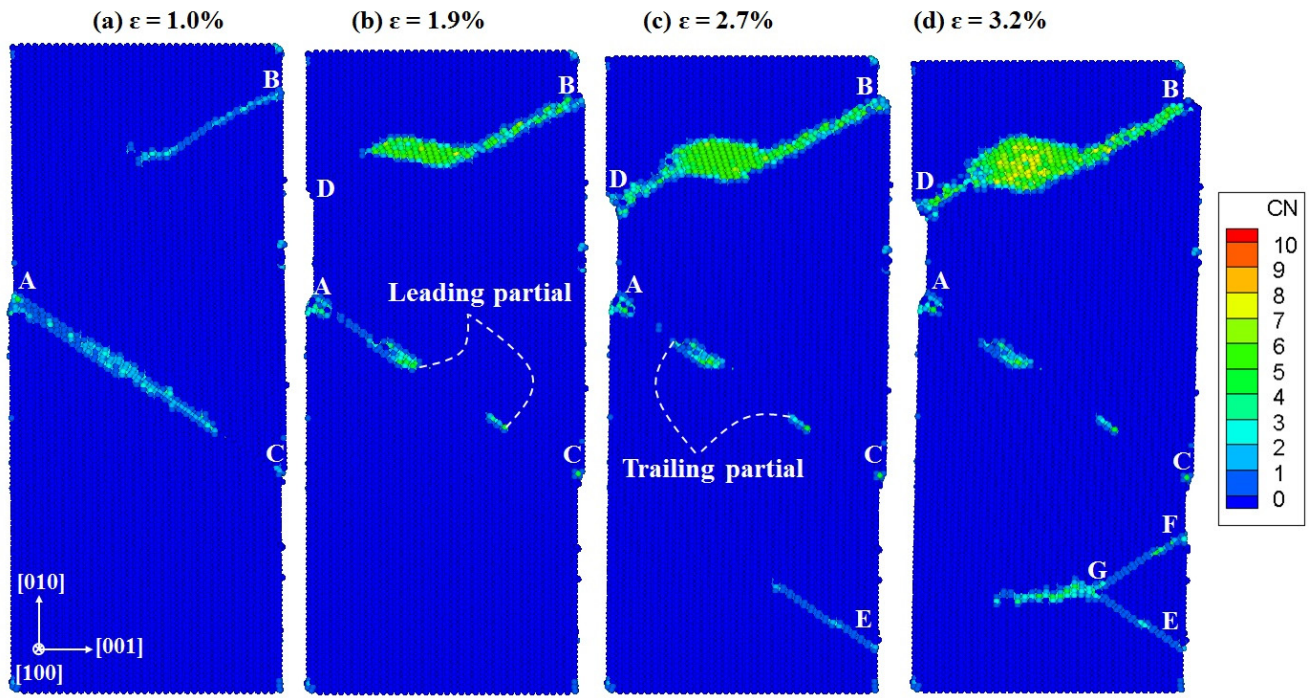

Fig. 10. Atomistic snapshots of deformed [010]-oriented YSTZ nanopillar at strains of (a) 1.0%; (b) 1.3%; (c) 1.7% and (d) 2.0%. Atoms are colored by coordination number (CN).